\documentclass[conference]{IEEEtran}
\IEEEoverridecommandlockouts
\usepackage{cite}
\usepackage{amsmath,amssymb,amsfonts}
\usepackage{algorithmic}
\usepackage{graphicx}
\usepackage{textcomp}
\usepackage{xcolor}
\usepackage{chngpage}
\def\BibTeX{{\rm B\kern-.05em{\sc i\kern-.025em b}\kern-.08em
    T\kern-.1667em\lower.7ex\hbox{E}\kern-.125emX}}
\begin{document}

\title{Data Science in an Agent-Based Simulation World}

\author{\IEEEauthorblockN{Satoshi Takahashi}
\IEEEauthorblockA{\textit{College of Science and Engineering} \\
\textit{Kanto Gakuin University}\\
Yokohama, Japan\\
satotaka@kanto-gakuin.ac.jp}
\and
\IEEEauthorblockN{Atushi Yoshikawa}
\IEEEauthorblockA{\textit{College of Science and Engineering} \\
\textit{Kanto Gakuin University}\\
Yokohama, Japan\\
atsuyoshi@kanto-gakuin.ac.jp}
}

\maketitle

\begin{abstract}
In data science education, the importance of learning to solve real-world problems has been argued. However, there are two issues with this approach: (1) it is very costly to prepare multiple real-world problems (using real data) according to the learning objectives, and (2) the learner must suddenly tackle complex real-world problems immediately after learning from a textbook using ideal data. To solve these issues, this paper proposes data science teaching material that uses agent-based simulation (ABS). The proposed teaching material consists of an ABS model and an ABS story. To solve issue 1, the scenario of the problem can be changed according to the learning objectives by setting the appropriate parameters of the ABS model. To solve issue 2, the difficulty level of the tasks can be adjusted by changing the description in the ABS story. We show that, by using this teaching material, the learner can simulate the typical tasks performed by a data scientist in a step-by-step manner (causal inference, data understanding, hypothesis building, data collection, data wrangling, data analysis, and hypothesis testing). The teaching material described in this paper focuses on causal inference as the learning objectives and infectious diseases as the model theme for ABS, but ABS is used as a model to reproduce many types of social phenomena, and its range of expression is extremely wide. Therefore, we expect that the proposed teaching material will inspire the construction of teaching material for various objectives in data science education.
\end{abstract}


\section{Introduction}
Data science is one of the fields that have attracted the most attention in recent years. As the popularity of data science has increased, so too has the need for educational methods, and many curricula and learning methods have been proposed and implemented. For example, the curriculum of “Data 8: The Foundations of Data Science” (University of California, Berkeley) covers three topics: inferential thinking, computational thinking, and relevance to the real world \cite{b1}. The curriculum teaches key concepts and related skills in computer programming and statistical inference through hands-on analysis of real-world datasets, including economic data, text data, geographic data, and social network data.

In classical data science education, lessons were based mainly on textbooks \cite{b2}. In such lessons, the learner learned how to apply ideal methods to ideal data and obtain ideal results \cite{b2}. More recently, data science education has emphasized the importance of real-world problems \cite{b3}. The aim is to enable learners to experience the tasks performed by a data scientist, including data collection, organization, analysis, and decision making, by working on real-world problems. In contrast to textbook problems, real-world problems require learners to start processing data from the data generation stage. Learners are required to read the background information about the data, deepen their understanding of the data, construct hypotheses, collect data according to the hypotheses, perform preprocessing, and analyze the data.

This real-world approach to learning presents two challenges. The first (issue 1) is that it is very costly to prepare multiple real-world problems (using real data) according to the learning objectives. For example, Hick et al. created data science case material using real data, but they pointed out that finding appropriate data incurs a major cost \cite{b3}. In addition, there are curricula in which learners work on data collection, but these curricula are limited to data that learners can access, such as data on familiar phenomena collected using sensors \cite{b4} or scraped from Web resources \cite{b5}.

The second issue (issue 2) is the gap between textbook problems and real-world problems. Textbook problems contain ideal data that satisfy the learning objectives. However, because of the difficulty of obtaining data on real-world problems, it is difficult to match the real-world data to the learning objectives, and it is difficult to facilitate learning by changing the learning objectives in a step-by-step manner. Therefore, immediately after learning from the textbook using ideal data, the learner suddenly needs to work with complex real-world data.

In this paper, we propose data science teaching material that uses agent-based simulation (ABS). ABS is a simulation method that uses autonomous decision makers called agents \cite{b6}. Objects modeled as agents include humans, animals, automobiles, organizations, and many others.

The proposed teaching material asks the learner to solve problems in a world constructed with ABS. The proposed teaching material comprises an ABS story and data generated by the ABS model. An ABS story comprises text that describes enough content for the reader (learner) to infer the connections between the elements of the ABS model. The learner reads the ABS story, understands the worldview of ABS, analyzes the data generated by the ABS model, and solves problems in the world of ABS. In this manner, the learner learns the essential concepts of data science.

The ABS world can be adjusted to the worldview that matches the learning objectives by changing the specifications and parameters of the ABS model. We exploit this characteristic to attempt to address issue 1. Furthermore, by changing the content of the ABS story, we can change the difficulty level at which the learner understands the ABS worldview. This is how we attempt to address issue 2.

In data science, it is necessary to learn multiple ways of thinking and many analytical methods. This paper focuses on the causal inference between these ways of thinking and analytical methods.

In the remainder of the paper, Section II describes causal inference, Section III describes the proposed teaching material, Section IV describes the details of the ABS model, Section V describes the ABS story and data generated by the ABS model, Section VI describes the educational approach that uses the proposed teaching material, Section VII discusses the use of the proposed teaching material, and Section VIII presents our conclusion.

\section{CAUSAL INFERENCE}
There are two approaches to causal inference: those of Pearl and Rubin \cite{b6,b7}. For our proposed teaching material, we adopted Pearl’s causal diagrams because of their visual clarity.

\begin{figure}[tbp]
\centerline{\includegraphics[width=0.8\linewidth]{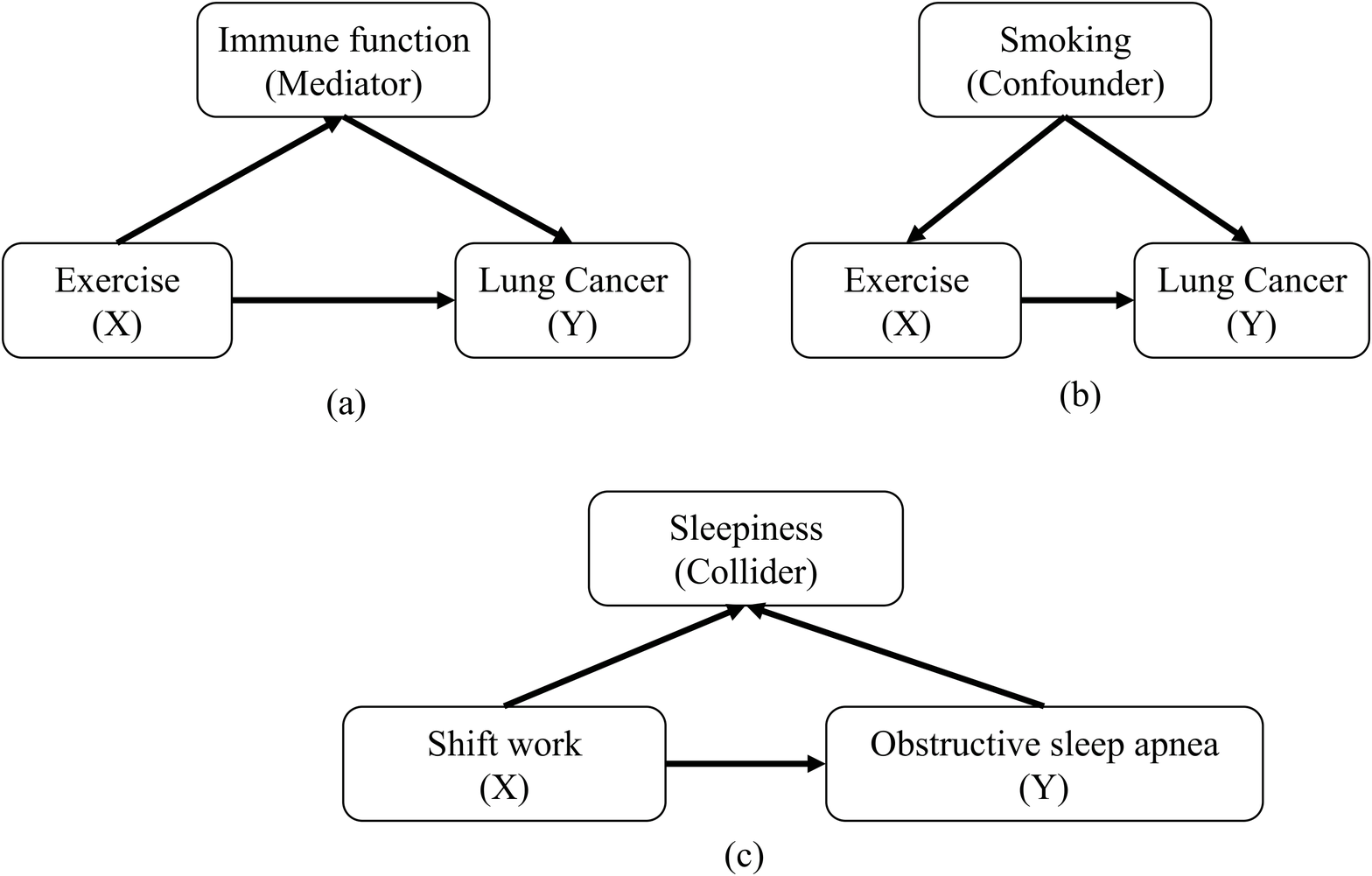}}
\caption{Sample causal diagrams (modified from \cite{b8}).}
\label{fig:1}
\end{figure}

The basic learning items in causal inference include the mediator, confounder, and collider \cite{b7,b8}. Fig. \ref{fig:1} shows their corresponding causal diagrams for an example in which there is causality from X to Y.

A mediator is a factor that is influenced by X and influences Y, as shown in Fig. \ref{fig:1}(a). For example, if we consider the effect of exercise (X) on lung cancer incidence (Y), immune function is the mediator. This is because it is thought that exercise improves immune function and immune function decreases the incidence of lung cancer. Therefore, if the data are separated by the level of immune function (the mediator), the direct effect of exercise (X) on lung cancer incidence (Y) may be estimated inaccurately.

A confounder is a factor that affects both X and Y in Fig. \ref{fig:1}(b). For example, if we consider the effect of exercise (X) on lung cancer incidence (Y), smoking is the confounder. This is because it is thought that the incidence of lung cancer increases and exercise declines in smokers. Therefore, separating the data by smoking (the confounder) may result in the discovery of a false relationship between exercise and lung cancer incidence.

A collider is a factor that is influenced by both X and Y in Fig. \ref{fig:1}(c). For example, when considering the effect of shift work (X) on obstructive sleep apnea (Y), sleepiness is the collider. This is because the presence of shift work and the presence of obstructive sleep apnea may increase sleepiness. Therefore, separating the data by the presence or absence of sleepiness (the collider) may result in the discovery of a false relationship between shift work and obstructive sleep apnea.

In causal inference, it is necessary to draw causal diagrams appropriately, discover the mediator, confounder, and collider, and process them appropriately. In the proposed teaching material, the learner first understands the world of ABS and then learns how to discover and process the mediator, confounder, and collider while drawing the causal diagrams.

\section{PROPOSED TEACHING MATERIAL}
\begin{figure}[tbp]
\centerline{\includegraphics[width=0.8\linewidth]{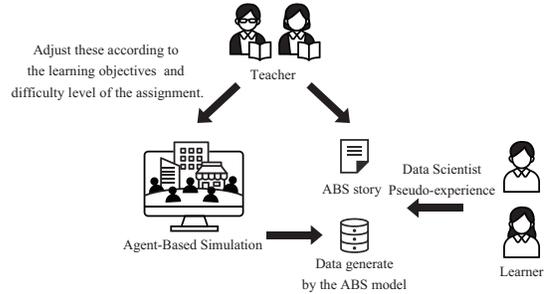}}
\caption{Overview of the proposed teaching material.}
\label{fig:2}
\end{figure}

In this paper, we propose data science teaching material in which scenarios and difficulty levels can be changed according to the learning objectives. The proposed teaching material consists of an ABS model and an ABS story (Fig. \ref{fig:2}).

We adopt infectious diseases as the modeling target of ABS. This is because we believe that learners have some knowledge of this subject as a consequence of the recent epidemic of COVID-19.

The learner is given the ABS story and the data generated by the ABS model. The learner then conducts a causal inference task to determine whether certain factors affect the probability of infection in the ABS world. The specific ABS parameters given to the learner are modified according to their learning objectives (i.e., whether they are learning to be a mediator, confounder, or collider).

\section{ODD PROTOCOL}
The ODD protocol is a method proposed by Grimm to describe ABS specifications and to ensure the reproducibility of ABS \cite{b9}. The ODD protocol contains Overview, Design concepts, and Details items. The Overview item contains three sub-items: Purpose, State variables and scales, and Process overview and scheduling. The Details item contains three sub-items: Initialization, Input, and Submodels. Because of space limitation, only some excerpts of the specifications are shown below.

\subsection{Overview}
This model is set in Kanazawa-ku, Yokohama City, Kanagawa Prefecture, where the author’s university is located. The reason for this is that the model is used in a class for students of the author’s university, so this setting makes it easy for them to construct their own worldview. The agents in this model are adults and children, and their whereabouts are their homes, workplaces, schools, restaurants, and hospitals. Adults and children may become infected with infectious diseases as they move between their homes, workplaces, schools, restaurants, and hospitals. The results of a 200-day simulation of the infection process are output and used as data.

\begin{figure}[tbp]
\centerline{\includegraphics[width=0.8\linewidth]{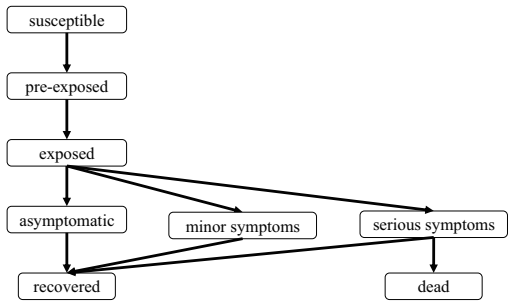}}
\caption{Infection state transition diagram.}
\label{fig:3}
\end{figure}

\subsubsection{Purpose}
The objectives of this model are
 \begin{itemize}
  \item to reproduce the process of spreading an infectious disease;
  \item to modify parameters to match the model causality to the learning objectives;
  \item to ensure that the model has a view of the world in which learners can naturally immerse themselves;
  \item but not to faithfully reproduce real phenomena.
 \end{itemize}

\subsubsection{State variables and scales}
The state variables of the model are shown in Table \ref{tab1}. To define the scale, one time step represents one hour.

\begin{table}[tbp]
\caption{STATE VARIABLES}
\begin{tabular}{|p{1.5cm}|p{6.5cm}|}
\hline
\textbf{\textit{Agent}}&\textbf{\textit{Variables}}\\ \hline
simulation manager&coordinates, adult list, child list, home list, restaurant list, workplace list, hospital list, probability of infection at home, probability of infection at work, probability of infection at school, probability of infection at restaurant\\ \hline
adult&name, home, workplace, height, weight, age, sex, accination status, probability of visiting restaurants on weekdays\\ \hline
child&name, home, school, height, weight, age, sex\\ \hline
home&name,latitude, longitude\\ \hline
workplace&name, latitude, longitude\\ \hline
school&name, latitude, longitude, online class status\\ \hline
restaurant&name, latitude, longitude, number of seats, short business hours status\\ \hline
hospital&name, latitude, longitude, number of beds\\ \hline
\end{tabular}
\label{tab1}
\end{table}

\subsubsection{Process overview and scheduling}
The ABS model simulates 200 days from 7 July 2022 to 22 January 2023. Adults and children are evaluated every hour for location and infection. They move to their homes, workplaces, schools, and restaurants according to the mobility algorithm (detailed in Section IV-C3). The mobility algorithm uses the concept of weekdays and holidays, where holidays include Saturdays and Sundays in addition to Japanese holidays.

According to the infection algorithm (details of which are presented in Section IV-C3), infection occurs stochastically.

\subsubsection{Design Concepts}
\subsubsection{Emergence}
Infectious diseases spread between adults and children under the influence of various factors. Adults and children may be infected when they spend time in the same place as an infected person. The probability of infection depends on vaccination status, age, and place (home, workplace, school, or restaurant).

\subsubsection{Interaction}
Adults and children are considered to be in contact if they spend time in the same place. If a person has contact with an infected person, he or she is probabilistically infected.

\subsubsection{Stochasticity}
The following elements are determined stochastically.
 \begin{itemize}
  \item Whether adults and children are infected when they come into contact with infected persons.
  \item Numbers of exposure days, asymptomatic days, days with minor symptoms, and days with severe symptoms.
  \item Transitions from exposed to asymptomatic, minor symptoms, and severe symptoms.
  \item Transitions from severe symptoms to dead or recovered.
  \item Whether to visit a restaurant.
  \item Who will visit a restaurant if there are more adults or children than available seats.
  \item Who will be admitted to a hospital if there are more adults or children than beds in the hospital.
 \end{itemize}

\subsubsection{Details}
Various parameters need to be set for initialization and input. Because of space limitations, this section focuses on the parameters that should be set according to the learning objectives (i.e., whether the confounder, mediator, or collider is to be learned).

\subsubsection{Initialization}
One adult is randomly selected and set to the exposed state. The other adults and children are set to the susceptible state.

Initially, there are 62,500 adults, 10,000 children, 50,000 homes, 500 workplaces, 100 restaurants, 7 hospitals, and 22 schools. The workplaces, homes, and restaurants are randomly placed at locations (latitudes and longitudes) in Kanazawa-ku, Yokohama City, Kanagawa Prefecture. The names and locations of hospitals and schools are set to those of the actual hospitals and schools in the area.

In addition to the above settings, there are other settings that depend on the scenario to be given to the learner. These are described in Section V.

\subsubsection{Input}
The external data are the location (latitude and longitude) of Kanazawa-ku, Yokohama City, Kanagawa Prefecture, and the names and locations of hospitals and schools located in the ward.

\subsubsection{Submodels}
\subsubsection{Migration algorithm}
Adults move from place to place according to the schedule presented in Tables \ref{tab2} and \ref{tab3}. Adults visit restaurants in a stochastic manner. On weekdays, they visit a restaurant near their workplace, and on weekends, they visit a restaurant near their home. If a restaurant is full, the second-nearest restaurant is selected. If that restaurant is full, the third- nearest is selected. If the third-nearest restaurant is full, they stays at workplace or home. Children also move from place to place according to the schedule presented in Tables II and III. On holidays, children follow adults who live with them when the adults go to a restaurant.

Restaurants have a “short business hours” status, which determines whether they have short hours. If a restaurant has short hours, adults and children are unable to visit the restaurant during the hours when it is closed. Schools have an “online class” status, which determines whether online classes are offered. If online classes are offered, children study at home.

\begin{table}[tbp]
\centering
\caption{WEEKDAY FLOW}
\begin{tabular}{|c|c|c|}
\hline
\textbf{\textit{Time}}&\textbf{\textit{Adult}}&\textbf{\textit{Child}}\\ \hline
0:00–9:00&home&home\\ \hline
9:00–12:00&workplace&school\\ \hline
12:00-13:00&restaurant or workplace&school\\ \hline
13:00–17:00&workplace&school\\ \hline
17:00–21:00&restaurant or home&home\\ \hline
21:00–24:00&home&home\\ \hline
\end{tabular}
\label{tab2}
\end{table}

\begin{table}[tbp]
\centering
\caption{HOLIDAY FLOW}
\begin{tabular}{|c|c|c|}
\hline
\textbf{\textit{Time}}&\textbf{\textit{Adult}}&\textbf{\textit{Child}}\\ \hline
0:00–12:00&home&home\\ \hline
12:00-13:00&restaurant or home&restaurant or home\\ \hline
13:00–17:00&workplace&school\\ \hline
17:00–21:00&restaurant or home&restaurant or home\\ \hline
21:00–24:00&home&home\\ \hline
\end{tabular}
\label{tab3}
\end{table}

\subsubsection{Infection algorithm}
Adults and children change their infection status according to the state transition diagram of Fig. \ref{fig:3}. The initial infection state is susceptible. When infected, the infection state changes to pre-exposed. On the following day, the state changes to exposed. After several more days, it transitions to either asymptomatic, minor symptoms, or severe symptoms in a stochastic manner. In the case of asymptomatic or minor symptoms, the patient transitions to the recovered state after several days. Several days later, a patient with severe symptoms probabilistically transitions to either dead or recovered.

\begin{eqnarray}
{Probability\ of\ infection}= \nonumber\\
1-(1-\alpha)^{\beta}\times(\frac{age}{10})^{3\times\gamma}\times(\frac{1}{10})^\delta
\label{eq}
\end{eqnarray}

Adults and children are infected stochastically if they spend time in the same place as an adult or child who is in the exposed, asymptomatic, minor symptoms, or severe symptoms state. Infection is determined on an hourly basis. The infection probability for each hour is determined by (1). Here, $\alpha$ is the probability of infection at each place for each home, school, workplace, and restaurant, $\beta$ is the number of infected people at that location, $\gamma$ is a parameter that determines how the infection probability increases with age, and $\delta$ is a parameter that determines the vaccination status; $\delta$ is 1 if the vaccine has been administered and 0 otherwise.

The overall probability of infection depends on the probability of infection at each place $\alpha$. If $\beta$ (the number of infected people at that location) is large, the probability of infection is high. If $\gamma$ is set to 0, there is no relationship between age and the probability of infection. However, if $\gamma$ is greater than 0, the probability of infection increases with age. If the person has been vaccinated, $\delta$ is set to 1 and the probability of infection is less than it is if $\delta$ is set to 0.

If a patient is infected but has minor symptoms, he or she stays at home and does not go to work, school, or restaurants. If the patient has severe symptoms, he or she is admitted to a hospital close to home. If the hospital is full, the second-nearest hospital is selected. If that hospital is full, the third- nearest is selected. If the third-nearest hospital is full, the patient stays at home.

\begin{figure}[tbp]
\centerline{\includegraphics[width=\linewidth]{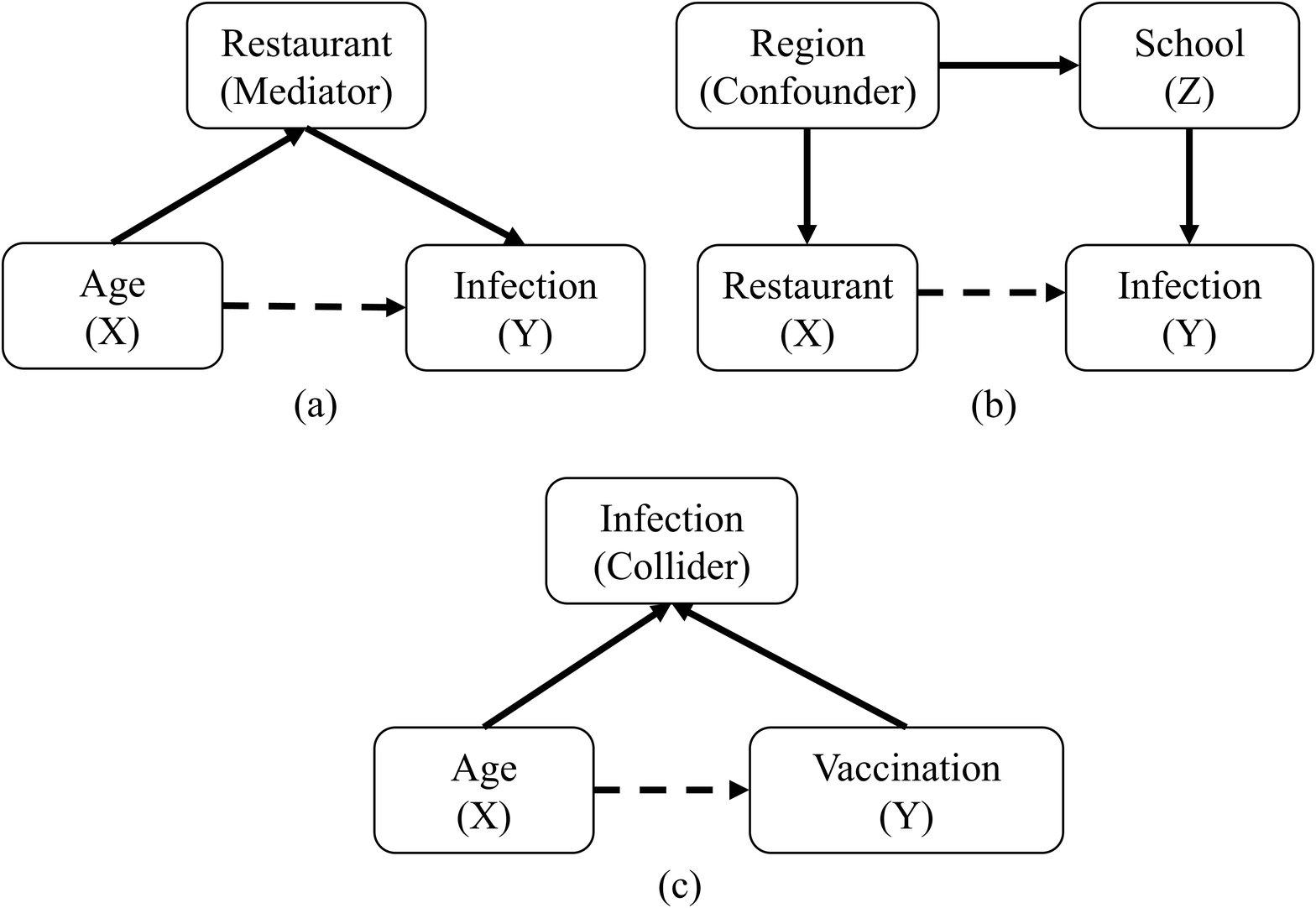}}
\caption{Causality diagram used in the teaching material, comprising (a) a mediator, (b) a confounder, and (c) a collider for the causality from X to Y. The dashed arrows do not represent real causal relationships.}
\label{fig:4}
\end{figure}

\section{MATERIAL GIVEN TO LEARNERS}
The learner is given an ABS story and the data generated by the ABS model.

\subsection{ABS Story}
In contrast to ODDs, an ABS story does not describe all the specifications of ABS. It describes enough for the reader (learner) to infer the connections between the elements of the ABS model. After reading these descriptions, the learner formulates hypotheses about cause-and-effect relationships between the elements. The ease of discovering causal relationships, and hence the level of the task, can be adjusted by changing the description in the ABS story.

The ABS story in the proposed teaching material is presented below and the method of level adjustment is described in Section VI.

You will work on causal inference of an infectious disease outbreak in a virtual world. The setting of the virtual world is Kanazawa-ku, Yokohama City, Kanagawa Prefecture. This is a closed world: people do not come and go from other areas. We will provide you with data for 200 days from 7 July 2022 to 22 January 2023, and you will be asked to analyze the factors that affect the infectious disease.

Two types of people appear in the virtual world: adults and children. Infectious diseases spread as adults and children move back and forth between their homes, workplaces, schools, and restaurants.

Each adult and child has a home where they live, work, and go to school. On holidays, both work and school are closed. They may visit restaurants for lunch or dinner. The extent to which they use restaurants varies from person to person. They choose restaurants that are as close as possible to their workplace or home.

Schools may offer online classes as a response to infectious diseases. There are three statuses of online classes. At status ZERO, all classes are face-to-face. At status ONE, there are face-to-face classes in the mornings and online classes in the afternoons. At status TWO, classes are entirely online.

Restaurants may also open for shorter hours as a measure to cope with infectious diseases. At status ZERO, restaurants are open all hours. At status ONE, they open only at noon. At status TWO, they remain entirely closed.

The changes in status when adults and children are infected with an infectious disease are shown in Fig. \ref{fig:3}. The initial state is the susceptible state. When an adult or child becomes infected, the infection state changes to pre-exposed. On the following day, it changes to exposed. After a few days, it transitions to asymptomatic, minor symptoms, or severe symptoms. In severe cases, death may occur after a few more days.

Adults and children may be exposed or asymptomatic; in both of these cases, they are unaware that they are infected and go about their daily routines as usual. In the case of minor symptoms, the patient may notice that he or she is infected and may miss work or school. In the case of severe symptoms, hospitalization is required. Patients try to be admitted to a hospital as close as possible to their home, but the number of beds is limited, so hospitalization is not always possible.

It is known that a person can be infected when an infected person is in the same space as him or her. The states in which a person can infect others are exposed, asymptomatic, minor symptoms, and severe symptoms. The probability of infection may vary, depending on various characteristics of the infecting person and the location where the infection occurs.

\subsection{Data Generated by ABS Model}
The data generated by the ABS model consist of 11 tables. Only some of the ABS data are included, namely the data that are necessary for learners to address the tasks given in the scenarios described in Section VI.

Table \ref{tab4} shows the details of the given data. The information tables for adults, children, homes, workplaces, schools, restaurants, and hospitals contain the relevant attribute information. The infection status tables for adults and children contain information on the infection status of individual adults and children, respectively. The place tables for adults and children contain the place for each hour of individual adults and children, respectively.

\begin{table}[tbp]
\caption{STATE VARIABLES}
\begin{tabular}{|p{3cm}|p{5cm}|}
\hline
\textbf{\textit{Table name}}&\textbf{\textit{Variables}}\\ \hline
adult information&name, home, vaccination, workplace, nearest restaurant to workplace, second-nearest restaurant to workplace, third-nearest restaurant to workplace, age, sex, height, weight\\ \hline
child information&name, home, school, age, sex, height, weight\\ \hline
home information&name, latitude, longitude, nearest hospital to home, second-nearest hospital to home, third-nearest hospital to home, nearest restaurant to home, second-nearest restaurant to home, third-nearest restaurant to home\\ \hline
workplace information&name, latitude, longitude, number of employees\\ \hline
restaurant  information&name, latitude, longitude, number of seats, short business hours status\\ \hline
school information&name, latitude, longitude, online class status\\ \hline
hospital information&name, latitude, longitude, number of beds\\ \hline
infection status of adult&date, infection status of each adult\\ \hline
infection status of child&date, infection status of each child\\ \hline
adult place&date, hour, place of each adult\\ \hline
ahild place&date, hour, place of each child\\ \hline
\end{tabular}
\label{tab4}
\end{table}

\section{Educational Methods Using Models}
By adjusting the parameters of the model, the instructor constructs scenarios that are tailored to the learning objectives. In addition, the ease of making causal hypotheses from the ABS stories can be adjusted by changing the wording of these stories. The learning objectives of the examples presented in this paper are the mediator, confounder, and collider, and scenarios for each of them are described below.

In each scenario, the learner is given a task to perform. While working on the task, the learner is expected to become aware of the existence of the mediator, confounder, and collider, and is expected to learn how to handle them appropriately. The scenarios are designed on the assumption that the learners have learned the basics of causal inference in advance.

\subsection{Mediator-Focused Scenarios}
Here we present a scenario that focuses on the mediator. In this scenario, the learner is asked the question, “Does age increase the likelihood of infection?” In this scenario, the probability of visiting a restaurant is set to decrease as the age of the adult increases, and $\gamma$  is not set to 0 in (1). This is a hypothetical scenario in which adults gradually become less likely to visit restaurants as they become older and the infection probability does not increase with age.

A causal diagram that relates age, restaurant visiting, and infection probability is shown in Fig. \ref{fig:4}(a). Because the probability of visiting a restaurant decreases with increasing age, an arrow is drawn from age to restaurant. In addition, because visiting a restaurant increases the chance of contact with others, an arrow is drawn from restaurant to infection. There is only a dashed arrow from age to infection.

Fig. \ref{fig:4}(a) shows that age does not directly increase the susceptibility to infection; instead, it increases it indirectly by mediating restaurant visits. That is, restaurant visits play a mediator role in the causal relationship between age and infection.

To adjust the level of the ABS story, we can replace “The extent to which they use restaurants varies from person to person” by “The frequency of restaurant visits varies with age.” Alternatively, we can insert “People who go out frequently are more likely to encounter an infected person” after “It is known that a person can be infected when an infected person is in the same space as him or her.” These added sentences encourage the learner to discover the causal path of the mediator in Fig. \ref{fig:4}(a).

\begin{figure}[tbp]
\centerline{\includegraphics[width=0.8\linewidth]{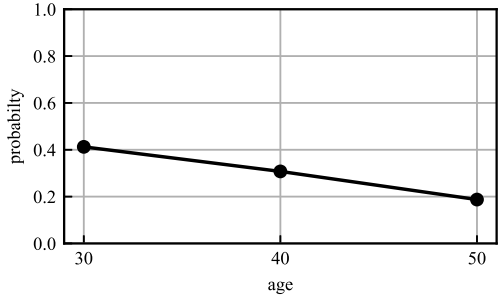}}
\caption{Probability of infection by age group.}
\label{fig:5}
\end{figure}

\begin{figure}[tbp]
\centerline{\includegraphics[width=0.8\linewidth]{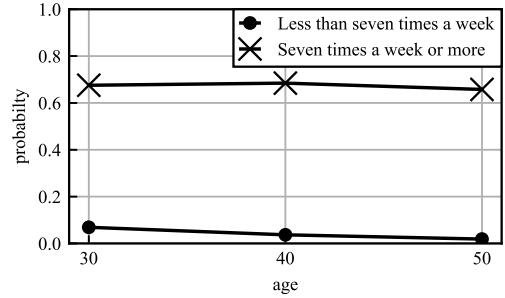}}
\caption{Probability of infection by age group and frequency of restaurant visits.}
\label{fig:6}
\end{figure}

When the probability of infection was calculated for each age group, the results are as shown in Fig. \ref{fig:5}, and it appears as if age decreases the probability of infection. However, when stratified by the number of times per week that restaurants are visited, the results are as shown in Fig. \ref{fig:6}. This indicates that the frequency of restaurant visits, rather than age, increases the probability of infection.

\subsection{Confounder-Focused Scenarios}
Here, we present a scenario that focuses on the confounder. In this scenario, the learner is asked the question, “Does limiting the opening hours of a restaurant reduce the probability of infection?” In this scenario, Kanazawa-ku is divided into several regions. For each region, the short business hours status of restaurants and the online class status of schools are positively correlated: For example, in one area, restaurants have high short business hour status and schools have high online class status. Furthermore, the $\alpha$  of restaurants (the probability of infection at restaurant) is set to 0 in (1). This is a hypothetical scenario in which each region has its own policy for short business hours at restaurants and online classes at schools: as the status of alertness for infectious diseases increases, the region increasingly forces restaurants to shorten business hours and schools to offer online classes. In addition, if adults or children visit a restaurant, the probability of infection is 0.

A causal diagram that relates region, restaurant visiting, school online class status, and infection probability is shown in Fig. \ref{fig:4}(b). Because each region has a different policy on restaurant hours and school online classes, arrows are drawn from region to restaurant and school. In addition, because a longer time spent at school leads to more opportunities for contact with others, an arrow is drawn from school to infection. Conversely, no arrow is drawn from restaurant to infection because the probability of infection in restaurants is set to 0.

Fig. \ref{fig:4}(b) shows that restaurant hours have no direct effect on infection. However, because the region influences both the business hours of restaurants and the online class status of schools, there is an apparent causal effect from restaurants to infection. The region plays the role of a confounder in the causal relationship between restaurant visits and infection.

To adjust the level of the ABS story, we can insert “It should be noted that each region seems to have its own measures to control infectious diseases in restaurants and schools” or “In addition, measures to prevent infectious diseases in restaurants and schools exist in each region. It appears that the more aware people are of the infectious disease crisis, the more likely they are to shorten the business hours of restaurants and face-to-face classes at schools” after “After a few days, it transitions to asymptomatic, minor symptoms, or severe symptoms. In severe cases, death may occur after a few more days.” These added sentences encourage the learner to discover the causal path of the confounder in Fig. \ref{fig:4}(b).

\begin{figure}[tbp]
\centerline{\includegraphics[width=0.8\linewidth]{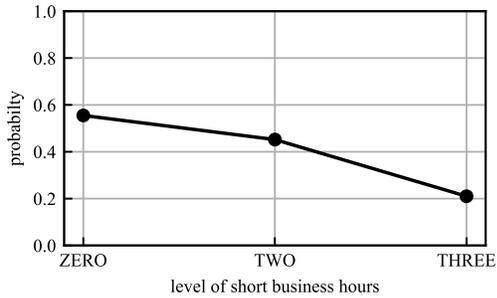}}
\caption{Probability of infection by status of short business hours.}
\label{fig:7}
\end{figure}

\begin{figure}[tbp]
\centerline{\includegraphics[width=0.8\linewidth]{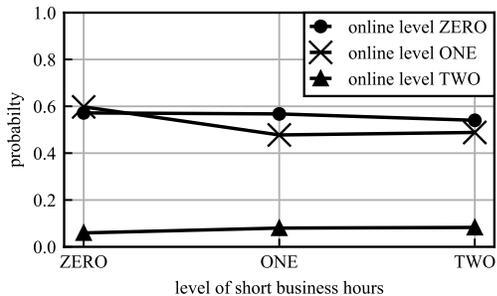}}
\caption{Probability of infection by status of short business hours and online class status.}
\label{fig:8}
\end{figure}

The probability of infection was calculated for each adult according to the status of short business hours of restaurants near the adult, as shown in Fig. \ref{fig:7}. It appears as if an increasing status of short business hours decreases the probability of infection. However, when the calculations combine the status of short business hours with the online class status of the children’s schools, the results are as shown in Fig. \ref{fig:8}. It can be observed that the status of short business hours does not affect the probability of infection.

\subsection{Collider-Focused Scenarios}
Here we present a scenario that focuses on the collider. We ask the learner, “We surveyed infected individuals and found that their vaccination rate appears to decrease as their age increases. Is this correct?” In this scenario, $\gamma$ (which determines how the infection probability increases with age) is set to a value greater than 0 in (1). In addition, the vaccination status $\delta$ of each adult is randomized. This is a hypothetical scenario in which the susceptibility to infection gradually increases with age.

A causal diagram that relates infection, vaccination rate, and age is shown in Fig. \ref{fig:4}(c). Because the infection rate increases with age, an arrow is drawn from age to infection. Another arrow is drawn from vaccination to infection because the probability of infection decreases with vaccination. However, there is only a dashed arrow between age and vaccination because there is no direct relationship.

Fig. \ref{fig:4}(c) shows that increasing age does not increase the vaccination rate. Restricting the sample to infected individuals decreases the proportion of vaccinated individuals and increases the age of the sample. That is, infection plays a collider role in the causal relationship between age and vaccination rate.

To adjust the level of the ABS story, we can insert “For example, the elderly may be more susceptible to infection” or “For example, vaccination may prevent infection” after “The probability of infection may vary, depending on various characteristics of the infecting person and the person being infected and the location where the infection occurs.” These added sentences encourage the learner to discover the collider causal path in Fig. \ref{fig:4}(c).

\begin{figure}[tbp]
\centerline{\includegraphics[width=0.8\linewidth]{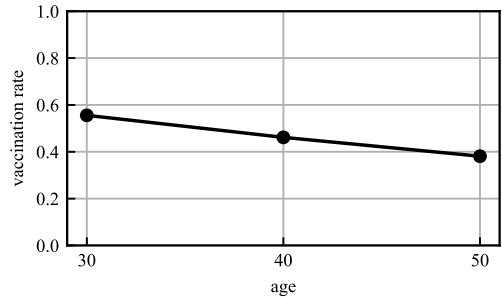}}
\caption{Vaccination rate by age group, restricted to infected persons.}
\label{fig:9}
\end{figure}

\begin{figure}[tbp]
\centerline{\includegraphics[width=0.8\linewidth]{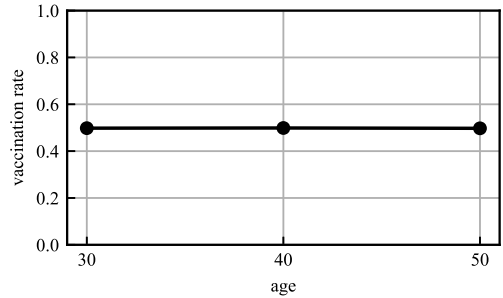}}
\caption{Vaccination rate by age group, not restricted to infected persons.}
\label{fig:10}
\end{figure}

When the vaccination rate was calculated only for infected persons and plotted for each age group, the results are as shown in Fig. \ref{fig:9}, and it appears as if the vaccination rate decreases with age. However, when the data are not divided between infected and susceptible persons, the results are as shown in Fig. \ref{fig:10}, indicating that there is no relationship between age and vaccination rate.

\section{DISCUSSION}
In this paper, we focus on the gap between education based on textbooks and education based on real-world problems in data science, and propose teaching material that can bridge this gap. We have proposed teaching material for data science in which the scenarios and difficulty levels can be changed according to the learning objectives.

The proposed teaching material consists of an ABS model and an ABS story. By changing the parameters of the ABS model, the instructor can construct an ABS world and generate data from them that satisfy the learning objectives. In addition, the instructor can change the ease of finding causal relationships and adjust the difficulty level of the task by changing the description in the ABS story. The learner can simulate a data scientist by reading the ABS story and analyzing the generated data.

We showed that, by appropriately adjusting the parameters of the ABS model, it is possible to construct scenarios that demonstrate the basic learning items of causal inference: the mediator, confounder, and collider. We also showed how to adjust the level of the ABS stories.

Simply providing a task does not guarantee that the learner will learn; appropriate facilitation by the instructor is necessary. However, the proposed teaching material allows the instructor to select a scenario that satisfies the learning objectives.

Typical tasks performed by a data scientist include data understanding, hypothesis building, data collection, data wrangling, data analysis, and hypothesis testing. In the following subsections, we explain how the proposed teaching material enables the learner to simulate these tasks.

\subsection{Data Understanding}
The learner gains a deeper understanding of the ABS model while reading the given ABS story and understanding the properties of the data.

The ABS story describes the behavioral rules for adults and children, the transition rules for infection, and the factors that affect them, but it does not include a causal diagram. Therefore, the learner needs to draw a causal diagram of the model world while reading and interpreting the ABS story.

For example, the learner would notice that infectious diseases are transmitted when people are in the same space from the sentence, “It is known that a person can be infected when an infected person is in the same space as him or her”. In addition, they would notice that the school’s online class status influences the chance of contact from the statement, “Schools may offer online classes as a response to infectious diseases.” Consequently, the learner would discover the causal effect that the school’s online class status influences the probability of infection.

\subsection{Hypothesis Building}
Not all causal relationships are explicitly described in the ABS stories. For example, the frequency of restaurant visits is only described in the sentence, “The extent to which they use restaurants varies from person to person.”

In response to this, the learner is required to formulate a hypothesis that there is some trend in the frequency of restaurant visits, and to link this hypothesis to various data and verify it. For example, the hypothesis is that the frequency of restaurant visits varies by region, sex, and age. This hypothesis-building work is embedded in every scenario.

\subsection{Data Collection and Data Wrangling}
To test the hypotheses constructed, the learner collects, connects, and processes the necessary data so that they can be used in the analysis.

Because the data are divided into various databases, the learner needs to extract the elements necessary for verification and then link these items together. For example, in the case of the confounder example, the learner needs to analyze information on adult infections, nearby restaurants, and the schools attended by children. This information is spread across an adult information table, a child information table, a school information table, and a restaurant information table. Therefore, it is necessary to link these tables together and compile the data.

In the example of infectious diseases used in this paper, data cleansing work is not required because errors and noise are not included in the data. However, prior research has identified the experience of data cleansing on real data as an essential component of data science education \cite{b10}.

By extending the proposed teaching material, data cleansing tasks can also be introduced. For example, we explicitly provide adult information as an adult information table. This is not realistic because, in general, information such as the height, weight, and address of all adults cannot be collected. Therefore, we add a sub-model to the ABS model in which this information is collected in the form of a questionnaire to be completed at the hospital when an adult has a disease. When completing this questionnaire, it is expected that adults will approximate their height using fractions, underestimate their weight, answer the number of days they have been experiencing symptoms dishonestly, and omit some items. This model replicates this behavior. It prevents the learner from directly referring to the adult information table, thus requiring the learner to collect survey information and eliminate errors and noise.

\subsection{Hypothesis Testing}
Using the processed data, the learner is required to test a hypothesis. For example, in the example mediator-focused scenario, it is possible to test whether a causal relationship exists between age and infection by adjusting for the frequency of restaurant visits.

\subsection{Summary}
The above discussion shows that the proposed teaching material allows the learner to simulate the typical tasks performed by a data scientist (data understanding, hypothesis building, data collection, data wrangling, data analysis, and hypothesis testing). The proposed teaching material is expected to bridge the gap between textbooks and real-world problems by changing the learning objectives in a step-by-step manner and gradually bringing the complexity of the model closer to that of a real-world problem.

The proposed teaching material described in this paper has some limitations. Most importantly, this paper only proposes teaching material; it does not report the results of demonstration experiments in actual educational settings. Such experiments are being conducted in the course “Data Science Application” at author’s University in the 2023 academic year. In the future, it will be necessary to consider ways to improve the system according to the results of these experiments. Nevertheless, the paper shows how the proposed teaching material can be modified by decomposing the educational objectives and creating scenarios that correspond to each of them.

\section{CONCLUSION}
The purpose of this paper is to propose data science teaching material in which the scenarios and difficulty levels can be changed according to the learning objectives. The teaching material consists of an ABS model and an ABS story. In this paper, as an example of the teaching material, we focused on causal inference as the learning objectives and infectious diseases as the model theme for ABS.

We have shown that, by appropriately adjusting the parameters of the ABS model, it is possible to generate scenarios that include the basic learning items of causal inference: the mediator, confounder, and collider. We have also shown that the difficulty level of the task can be changed by adjusting the wording of the ABS story. Furthermore, we have shown that the learner can simulate the typical tasks performed by a data scientist (data understanding, hypothesis building, data collection, data wrangling, data analysis, and hypothesis testing) by using this material.

Although we have used infectious diseases as an example, ABS has been used as a model to reproduce many types of social phenomena, and its expressiveness is very high. Therefore, it is expected that the proposed teaching material will lead to the construction of teaching material for various objectives in data science education.

\section*{Acknowledgment}
We thank Edanz (https://jp.edanz.com/ac) for editing a draft of this manuscript.


\begin{thebibliography}{00}
\bibitem{b1} ``Data 8: The Foundations of Data Science.'' http://www.data8.org/ (accessed May 11, 2023).
\bibitem{b2} C. J. Wild and M. Pfannkuch, ``Statistical thinking in empirical enquiry,'' Int. Stat. Rev., vol. 67, no. 3, pp. 223--248, 1999.
\bibitem{b3} S. C. Hicks and R. A. Irizarry, ``A guide to teaching data science,'' Am. Stat., vol. 72, no. 4, pp. 382--391, 2018.
\bibitem{b4} T. Lechler, M. Sjarov, and J. Franke, ``Data farming in production systems — A review on potentials, challenges and exemplary applications,'' Procedia CIRP, vol. 96, pp. 230--235, 2021.
\bibitem{b5} M. Dogucu and M. Çetinkaya-Rundel, ``Web scraping in the statistics and data science curriculum: Challenges and opportunities,'' J. Stat. Data Sci. Educ., vol. 29(sup1), pp. S112–S122, 2021.
\bibitem{b6} R. Axelrod, The Complexity of Cooperation: Agent-Based Models of Competition and Collaboration. Princeton University Press, 1997.
\bibitem{b7} J. Pearl, ``Causal diagrams for empirical research,'' Biometrika, vol. 82, no. 4, pp. 669--688, 1995.
\bibitem{b8} D. J. Lederer et al., ``Control of Confounding and Reporting of Results in Causal Inference Studies. Guidance for Authors from Editors of Respiratory, Sleep, and Critical Care Journals,'' Ann. Am. Thorac. Soc., vol. 16, no. 1, pp. 22--28, 2019.
\bibitem{b9} V. Grimm et al., ``A standard protocol for describing individual-based and agent-based models,'' Ecol. Model., vol. 198, no. 1–2, pp. 115--126, 2006.
\bibitem{b10} S. Kross, R. D. Peng, B. S. Caffo, I. Gooding, and J. T. Leek, ``The democratization of data science education,'' Am. Stat., vol. 74, no. 1, pp. 1--7, 2020.

\end{thebibliography}
\end{document}